# Analysis of periodic Mo/Si multilayers: influence of the Mo thickness

H. Maury[1], J.-M. André[1], K. Le Guen[1], N. Mahne[2], A. Giglia[2], S. Nannarone[2], F. Bridou[3], F. Delmotte[3], P. Jonnard[1]

(1) UPMC Univ Paris 06, CNRS - UMR 7614, Laboratoire de Chimie Physique - Matière et Rayonnement, 11 rue Pierre et Marie Curie, F-75231 Paris cedex 05, France

(2) TASC-INFM National Laboratory, S.S. 14, Km 163.5 in Area Science Park, I-34012 Basovizza (Trieste), Italy

(3) Laboratoire Charles Fabry de l'Institut d'Optique, CNRS, Univ Paris-Sud, Campus Polytechnique, RD128, F-91127 Palaiseau cedex, France

*Corresponding author* : Dr. P. Jonnard, Laboratoire de Chimie Physique, 11 rue Pierre et Marie Curie, F-75231 Paris cedex 05, France
Tel : 33 1 44 27 63 03        Fax : 33 1 44 27 62 26        jonnard@ccr.jussieu.fr

A set of Mo/Si periodic multilayers is studied by non destructive analysis methods. The thickness of the Si layers is 5 nm while the thickness of the Mo layers changes from one multilayer to another, from 2 to 4 nm. This enables us to probe the effect of the transition between the amorphous to crystalline state of the Mo layers near the interfaces with Si on the optical performances of the multilayers. This transition results in the variation of the refractive index (density variation) of the Mo layers, as observed by x-ray reflectivity (XRR) at a wavelength of 0.154 nm. Combining x-ray emission spectroscopy and XRR, the parameters (composition, thickness and roughness) of the interfacial layers formed by the interaction between the Mo and Si layers are determined. However, these parameters do not evolve significantly as a function of the Mo thickness. It is observed by diffuse scattering at 1.33 nm that the lateral correlation length of the roughness strongly decreases when the Mo thickness goes from 2 to 3 nm. This is due to the development of Mo crystallites parallel to the multilayer surface.

*Corresponding author*: Dr. Philippe JONNARD
Laboratoire de Chimie Physique, 11 rue Pierre et Marie Curie
F-75231 Paris Cedex 05, France
Tel: 33 1 44 27 63 03      Fax: 33 1 44 27 62 26
e-mail: philippe.jonnard@upmc.fr

# 1. INTRODUCTION

The development of efficient periodic multilayers can benefit from advanced analysis techniques that can characterize these complex structures and help in understanding the phenomena taking place at their interfaces. Indeed, it is important to obtain a relevant description of a multilayer, i.e. to know the thickness and roughness of all the various layers, the composition, thickness and roughness of the interfacial zones, if any, the correlation lengths of the roughness, etc. These informations enable the improvement of the preparation of the multilayers and lead to the choice of a strategy to adapt the multilayer for long-time operation or hot environment.

In this paper, we study a series of periodic Mo/Si multilayers by following the methodology developed in previous papers [1-6], combining non-destructive techniques, X-ray emission spectroscopy (XES), x-ray reflectivity (XRR) and diffuse scattering measurements. Using XES, we deduce the chemical composition of the interfacial zones and estimate their respective thickness. Using XRR, we determine the thicknesses, the optical indices of all the layers and the *rms* height of their roughness. Using diffuse scattering, the lateral correlation length of the roughness can be estimated. This approach is presently applied in the case of a set of multilayers having amorphous Si layers of the same thickness, but Mo layers of different thickness. Thus, the study is performed as a function of the Mo thickness to probe the amorphous-crystalline transition around 3 nm.

## 2. EXPERIMENTAL DETAILS

### 2.1. Sample preparation

The Mo/Si multilayers are prepared by magnetron sputtering using an apparatus described elsewhere [7]. Here, we briefly recall the main characteristics of the deposition process. Argon at a pressure of 2 mTorr (1 Torr = 133.3 Pa) was used in the deposition chamber. The plasma discharges were obtained with a RF power of 150 W for Si targets and a DC current of 0.19 A for the Mo target. Samples are deposited on Si polished wafers. The number of bilayers is 40. Three samples with different Mo layers thicknesses have been fabricated, ranging around the transition thickness from the amorphous to the crystalline state, *i.e.* 2, 3 and 4 nm. In all samples, the thickness of the Si layers is 5 nm. The thickness of the different layers as well as the name given to each sample are listed in Table I.

*Table I: Name of the Mo/Si multilayers and expected thickness of the Mo layers from single layer calibration of the deposition process. All the Si layers are 5 nm thick.*

| Sample name | Mo thickness (nm) |
|---|---|
| **Mo-2** | 2.01 |
| **Mo-3** | 2.99 |
| **Mo-4** | 4.01 |

For XES, the reference samples are a 60 nm thick amorphous Si film deposited on a GaAs substrate and high purity silicide powders: $MoSi_2$ (Aldrich, purity 99%) and $Mo_5Si_3$ (Alfa Aesar, purity 99.5%).

## 2.2. X-ray emission spectroscopy

The Si Kβ emission coming from the silicon atoms present in the Mo/Si multilayers has been analysed. It corresponds to the 3p-1s transition and describes the occupied valence states having the Si 3p character. This emission is very sensitive to the physico-chemical state of the silicon atoms [8-9]. The X-ray analysis was performed in a high-resolution bent-crystal soft X-ray spectrometer [10] using an InSb (111) crystal at the first diffraction order.

The Si 1s core holes are created by an electron beam coming from a Pierce gun. The energy of the incident electrons was sufficiently low so that the electrons cannot reach the silicon substrate. Then, no signal from the substrate can interfere with that of the multilayer. The current density impinging on the sample is sufficiently low (some tenths of $mA/cm^2$) so that any evolution of the samples under the electron beam is avoided. It was checked during acquisition that the shape and the intensity of the emission do not vary.

## 2.3. Grazing incidence X-ray reflectivity

The experimental reflectivity curves are obtained by means of a reflectometer working with Cu Kα radiation of 0.154 nm wavelength. The Cu Kα radiation is selected by means of a graphite monochromator in front of the detector. The reflectivity curve is obtained by varying the grazing incidence angle while tracking the reflected beam (θ-2θ scan). The maximum angular amplitude is θ = 6° with an angular accuracy better than 5/1000° [11].

**2.4. Soft x-ray reflectivity and diffuse scattering measurements**

The reflectivity measurements at 0.133 nm were taken in $\theta$-$2\theta$ mode at the BEAR beamline [12] at Elettra. The reflectometer features an overall accuracy on the absolute reflectivity of ≈1%. The diffusion measurements are obtained in the transverse scan mode at 0.133 nm. The scattered intensity is measured at a fixed position of the detector while the sample is rotated [13]. The $\theta$ goniometer angular resolution was 0.01°. Impinging and reflected intensities were measured by an IRD SXUV100 solid state diode within two separate runs; incident intensities were monitored by a Au mesh inserted in the beam path whose drain current was used for normalisation.

## 3. RESULTS

**3.1. X-ray emission spectroscopy and X-ray reflectivity measurements**

In order to determine the physico-chemical state of the Si atoms in a given Mo/Si multilayer, their Si 3p spectral densities are compared to those of reference materials: a-Si, $MoSi_2$ and $Mo_5Si_3$. Such a comparison is shown in Figure 1 for Mo-2 as an example. The spectra are normalized with respect to their maximum and a linear background has been removed. If no interaction takes place between the Si and Mo layers, the spectrum of a-Si should be observed. The presence of two silicides at the interfaces is demonstrated by the broadening of the multilayer spectrum with respect to the amorphous Si spectrum, toward both the low and high photon energies where the maxima of $MoSi_2$ and $Mo_5Si_3$ are present.

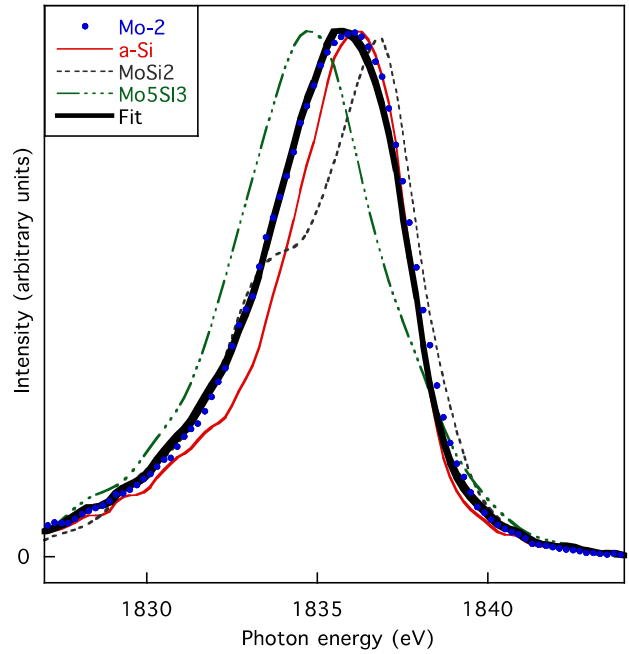

*Figure 1: Si Kβ soft x-ray emission spectra of the sample Mo-2 and the various references and fit of the multilayer spectrum by a weighted sum of the reference spectra.*

To determine the contribution of the silicides in the spectra and then estimate the interphase thickness, the multilayer spectra are fitted as a weighted sum of the reference spectra, as now routinely done in XES for the study of buried interfaces [5]. An example of fit is presented in the Figure 1 for the Mo-2 sample. The same operation is performed for the Mo-3 and Mo-4 samples that present Si 3p spectral densities close to that of Mo-2. The calculation of the interphase thickness is based on the model presented in Ref. [5,14] in which no distinction is made between the Mo-on-Si and Si-on-Mo interfaces. Because the multilayer spectra are quite similar, the interface (silicide) thickness is deduced to be almost the same, $0.8 - 0.9$ nm depending on the samples (see Table II), within the experimental uncertainty (±0.2 nm). These values are in the range of the thicknesses presented in the literature for non-annealed samples, see Ref. [15-17] for example, and determined from transmission electron microscopy images, XRR or XES experiments. The interfacial thickness varies according to the preparation conditions, the mechanical stress within the multilayer, the considered interface (in fact the Mo-on-Si interface is wider than the Si-on-Mo one), …

*Table II: Thickness and roughness of the various layers of the Mo/Si multilayers deduced from XES and from the fit of the reflectivity curves obtained at 0.154 nm. IL1 means interlayer at the Mo-on-Si interface. IL2 means interlayer at the Si-on-Mo interface. $\delta$ is the unit decrement of the real part of the refractive index.*

| Sample name | Interlayer thickness (nm) deduced from XES | Mo thickness (nm) <br> IL1 thickness (nm) <br> Si thickness (nm) <br> IL2 thickness (nm) | $\delta Mo$ (x$10^{-6}$) <br> $\delta IL1$ (x$10^{-6}$) <br> $\delta Si$ (x$10^{-6}$) <br> $\delta IL2$ (x$10^{-6}$) | Mo roughness (nm) <br> Il1 roughness (nm) <br> Si roughness (nm) <br> Il2 roughness (nm) |
|---|---|---|---|---|
| **Mo-2** | $0.9 \pm 0.2$ | 1.46 <br> 0.85 <br> 3.18 <br> 0.85 | 24.4 <br> 14.5 <br> 6.9 <br> 15.7 | 0.4 <br> 0.3 <br> 0.2 <br> 0.2 |
| **Mo-3** | $0.8 \pm 0.2$ | 2.51 <br> 0.80 <br> 3.26 <br> 0.87 | 26.4 <br> 16.2 <br> 6.8 <br> 12.9 | 0.3 <br> 0.4 <br> 0.1 <br> 0.4 |
| **Mo-4** | $0.8 \pm 0.2$ | 3.31 <br> 0.85 <br> 3.35 <br> 0.92 | 26.8 <br> 14.9 <br> 6.8 <br> 12.3 | 0.2 <br> 0.4 <br> 0.2 <br> 0.3 |

We fitted the grazing incidence reflectivity measurements by a four-layer model in a period: Mo/interlayer/Si/interlayer by taking into account the XES results (thickness of the interlayer, relative proportion of $MoSi_2$ and $Mo_5Si_3$ within the interlayer) as input parameters. The fit of the reflectivity curves is made using a trial and error method. It allows the determination of the thickness, the interfacial roughness, and at the source wavelength, the complex index n = 1 - δ - iβ (δ is the unit decrement of the refractive index and β is the extinction coefficient) for each of the successively deposited films on the substrate [18].

The composition of the interfaces (relative proportions of the silicides) deduced from XES is taken into account through δ. Thus, the unit decrement of the interfacial layer corresponds to a weighted sum of the unit decrements of the silicides. As an example, the Figure 2 shows the reflectivity curve of Mo-2 and its fit. The seven Bragg peaks are well reproduced as well as their relative intensity and the background between the peaks. The structural parameters of the three studied multilayers are indicated in the Table II. These are the thickness and roughness of each of the four layers. It is seen that the period of the stack is smaller than the expected one. This well-known contraction period is due to the formation of silicides that have a density greater than the corresponding mixing of Mo and Si. The roughness's have all the same magnitude in all samples. The main point is the change of the δ value of the Mo layer as a function of the thickness. Since the δ value is proportional to the density of the layer, δ(Mo) is higher for the crystalline state ($27 \cdot 10^{-6}$) than for the amorphous state (around $24 \cdot 10^{-6}$). We observe that the δ value of the Mo layer for the Mo-2 sample is that of amorphous Mo while for the Mo-3 and Mo-4 samples it is that of crystalline Mo.

The δ value is proportional to the density of the material, so the increase of the Mo density should lead to little more differences between Mo and Si indices and then increase reflectivity. It depends in fact of the choice of the working wavelength. For example, in our case, $\delta_{Mo} = 24.4 \cdot 10^{-6}$ corresponds to a density $d_1$ of 85% of the massive one d, and $\delta_{Mo} = 26.8\ 10^{-6}$ to $d_2 = 93\%$ of d. In normal incidence, with a two layers period Mo (23.1 nm)/Si (40.6 nm), the calculated reflectivity, corresponding to a Bragg peek at the wavelength of 12.15 nm, would be 0.258 for $d_1$ and 0.289 for $d_2$.

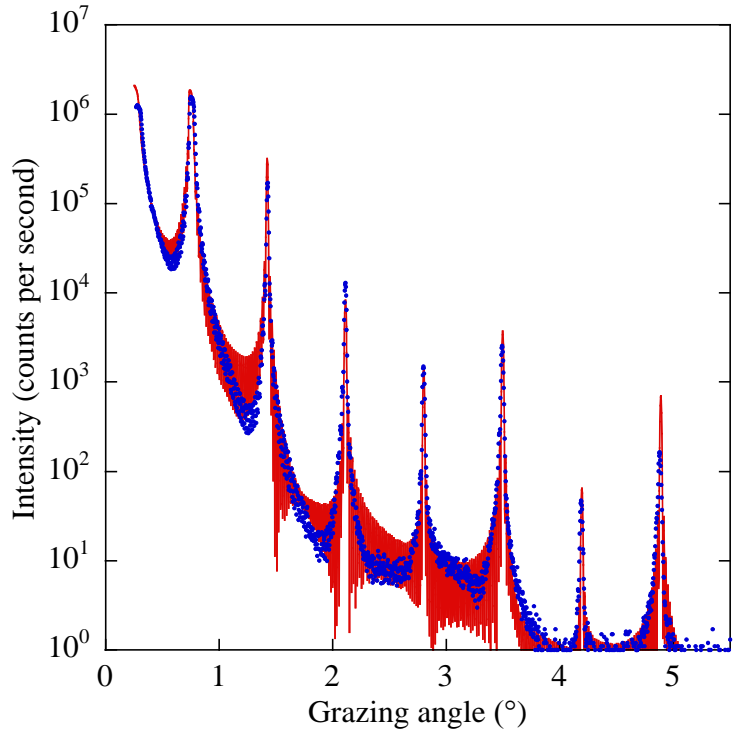

*Figure 2: X-ray reflectivity of the Mo-2 sample obtained at 0.154 nm with its fit by a model with four layers in a period.*

The reflectivity measurements in the soft x-ray range are performed at 1.33 nm (Cu L$\alpha$ emission). At this wavelength the first two Bragg peaks can be analysed. The structural parameters determined at 0.154 nm are used to simulate the reflectivity measurements; the $\delta$ values are extrapolated at the wavelength of 1.33 nm from the measurements at 0.154 nm. The comparison between this simulation and the experimental reflectivity is shown in the Figure 3 for the Mo-2 sample. Let us emphasize that there is a rather good agreement between experimental reflectivity measured in the soft x-ray and the simulation deduced from the values extrapolated from the hard x-ray measurements. Such extrapolation requires great care because the use of roughness parameters deduced from measurements at a given photon energy is restricted to non-obvious conditions in terms of photon energy and autocorrelation function [19-21]. This agreement confirms the relevant description of the Mo/Si multilayers obtained from the XES and XRR measurements.

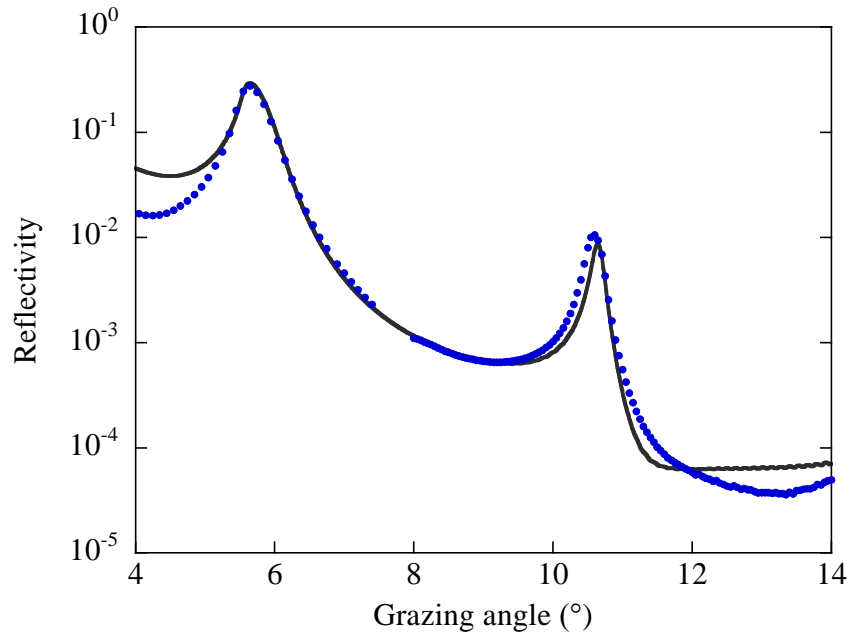

*Figure 3: Reflectivity of the Mo-2 multilayer at 1.33 nm (dots) compared to the simulation (line) deduced from XES and XRR at 0.154 nm.*

### 3.2. Diffuse scattering measurements

The diffusion measurements were made in the so-called transverse scan mode with a *s*-polarized radiation of 1.33 nm (930 eV). The samples have been rotated by $\pm 7°$ with respect to the specular direction. The measurements have been performed around the first and second Bragg peaks and are presented in the Figure 4. The intensities are normalized to unity. It is observed that the shape of the curves varies from that of Mo-2, specular peak on a bell shape curve, and those of Mo-3 and Mo-4, specular peak on a rather flat background.

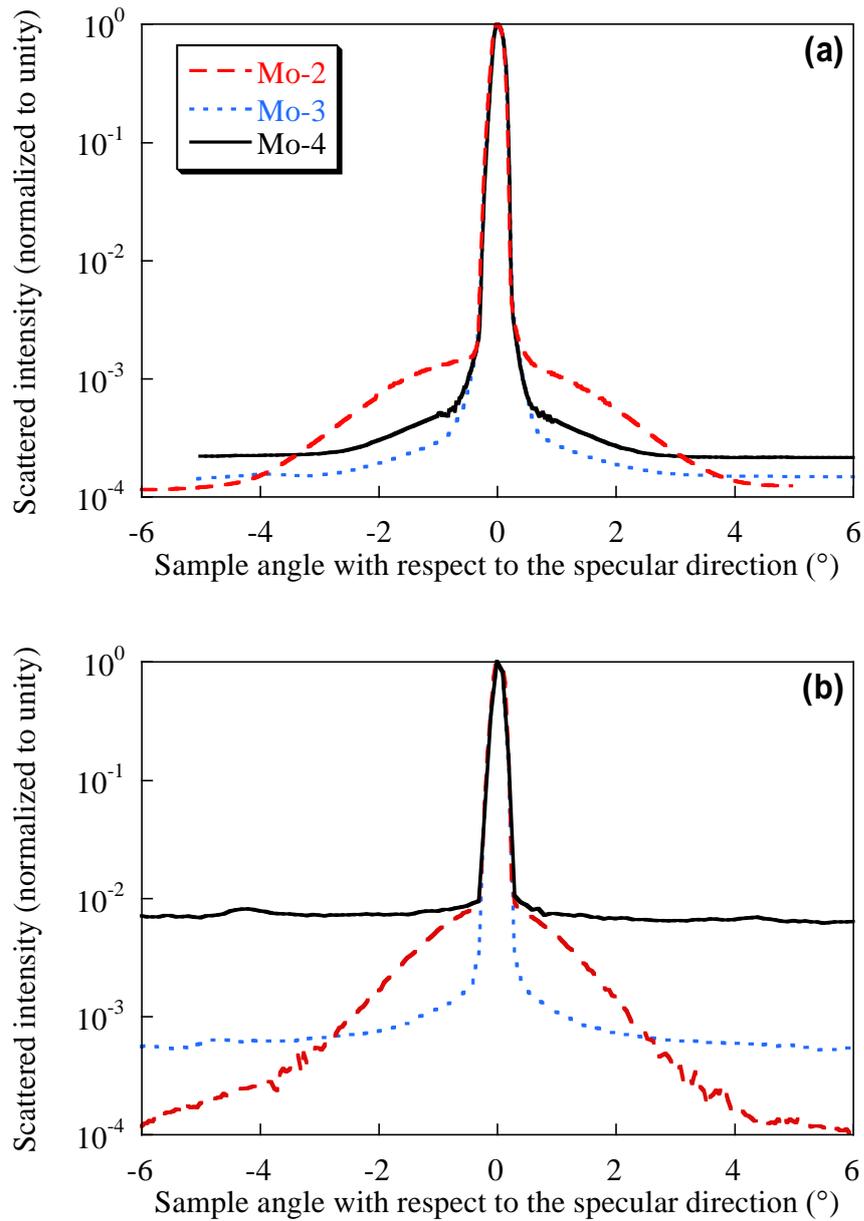

*Figure 4: Transverse scans at 1.33 nm around the first (a) and second (b) Bragg peaks for multilayers Mo-2 (dashed line), Mo-3 (dotted line) and Mo-4 (solid line).*

## 4. DISCUSSION

It is clear from the XES and XRR results that the Mo/Si multilayers are found to be a structure with four layers in a period: Mo/silicides/Si/silicides. This is well established for this system [22-24]. As a function of the thickness of the Mo layers, we observe no relevant change of the thickness, roughness and composition of the interlayers (Cf. Table II). The main change concerns the structure of the Mo layers that are amorphous when 2 nm thick and crystalline when 3 or 4 nm thick. This is in agreement with previous studies dealing with the structure of the metal layers within multilayers [23,25-26]. For example, by combining XRR, x-ray diffraction and transmission electron microscopy (TEM), Bajt et al. [26] have clearly evidenced the nucleation of the Mo crystallites when the layer thickness is greater than 2 nm. We deduce from our methodology that the various samples have almost the same *rms* roughness's, 0.2 – 0.4 nm: this is because amorphous or polycrystalline Mo generate layer roughness of the same magnitude. It is only in the small thickness range where the Mo crystallites form that the roughness increases [26].

The agreement on the total interlayer thickness is good between the value obtained with our methodology (around 1.7 nm) and the one determined from TEM images (around 1.8 nm) [27]. However, only the presence of $MoSi_2$ is deduced from the TEM analysis [26,27], whereas it is necessary to take into account both $MoSi_2$ and $Mo_5Si_3$ silicides to fit the XES spectra of the multilayers. We think that this discrepancy is due to the energy brought to the samples during their thinning by an ion beam needed for the TEM observation, whereas XES is a non-destructive technique. This energy promotes the formation of the most stable silicide, i.e. $MoSi_2$, leading the evolution of the original interlayer stoichiometry. The same compound, $MoSi_2$, is observed by x-ray photoelectron spectroscopy when the multilayers are etched by an ion beam in order to make profilometry measurements [28].

Despite *rms* roughness's are almost the same (see Table II), the diffuse scattering curves present drastically different shapes with the Mo thickness varying.

This puts into evidence that the lateral distribution of the roughness varies according to the structure of the Mo layers. Thus, we have fitted the diffusion curves in order to determine the lateral and vertical correlation lengths of the roughness, noted $\xi$ and $L$ respectively, using the IMD simulation software [29].

In these simulations, the intensity of the Bragg peak is not taken into account. In fact, the simulated curves are not very sensitive to the L values : we have varied L between 0 and 30 nm and did not found relevant changes. So the vertical correlation length has been arbitrary fixed at 10 nm and only the value of lateral correlation length was changed until a good agreement was found. As an example, we present the result of this procedure for the Mo-2 sample, in Figure 5 around the first Bragg peak and in Figure 6 around the second Bragg peak. For the first Bragg peak, the ends of the curves, ranges between 0 and 2° and between 10 and 12°, are not well reproduced. This is because the experiment is performed with a detection system having a counting dynamics limited to about 6 orders of magnitude, whereas the simulation assumes detection with an infinite dynamics.

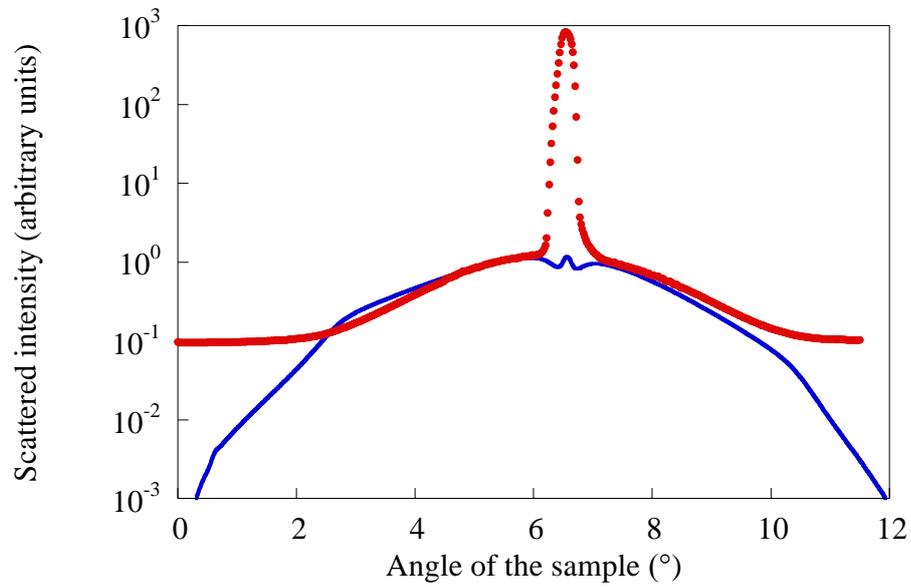

*Figure 5: Transverse scan (dots) at 1.33 nm around the first Bragg peak of the Mo-2 sample and simulation (line) with $\xi = 20$ nm and $L = 10$ nm.*

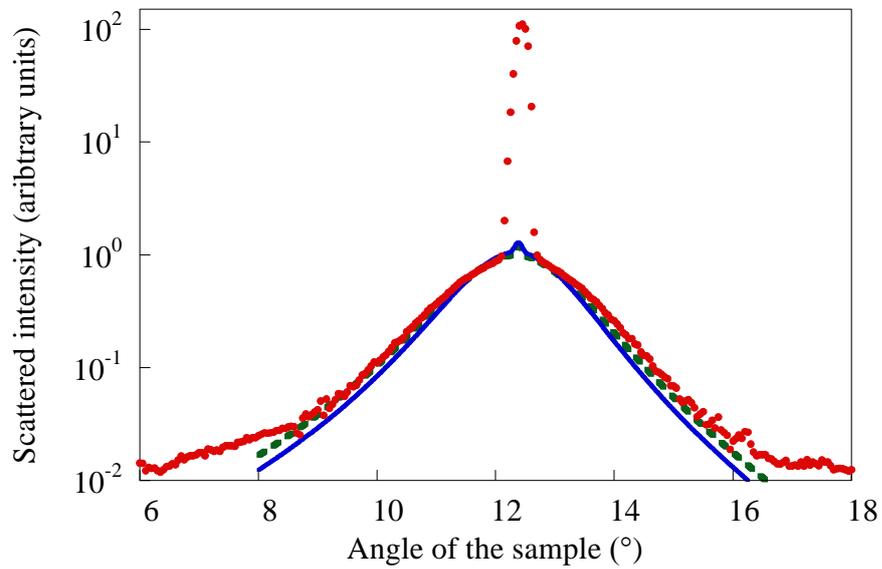

*Figure 6: Transverse scan (dots) at 1.33 nm around the second Bragg peak of the Mo-2 sample and simulation with $\xi = 20$ nm (solid line) et $\xi = 18$ nm (dotted line) and $L = 10$ nm.*

We present in Table III the lateral correlation length $\xi$ of the roughness, obtained around the two studied Bragg peaks and for the three samples. In each case, we indicate the range of the values that give a good agreement between the experiment and the simulation. For Mo-3 and Mo-4, the curves have been fitted only at the second Bragg peak, because we observe a small sensitivity of the shape of the curves to $\xi$ value around the first Bragg peak.

*Table III: Lateral correlation lengths $\xi$ of the roughness (nm) of the Mo/Si multilayers.*

| Sample name | first Bragg peak | second Bragg peak |
|:---:|:---:|:---:|
| **Mo-2** | 20 | 18 – 20 |
| **Mo-3** | - | 7 – 9 |
| **Mo-4** | - | 5 |

For the thinnest Mo layer, the largest lateral correlation length $\xi$ is estimated, *i.e.* about 18 nm for the Mo-2 sample. As soon as the Mo film becomes crystalline, *i.e.* when its thickness is equal to or larger than 3 nm, the correlation length decreases. Then Mo-3 and Mo-4 samples give $\xi$ values of the same magnitude. This means that the replication frequency of the roughness profile increases at the interfaces. This is due to the development of the Mo crystallites parallely to directions of the surface substrate, as shown by Bajt et al. [26].

## 5. CONCLUSION

The main effect of the Mo thickness on the structure of Mo/Si multilayers comes from the transition between the amorphous to crystalline state of the Mo layers. This transition does not affect in a significant way the thickness of the interfacial layers and their *rms* roughness, but has a large influence on the value of the refraction index of the Mo layers and on the lateral correlation length $\xi$ of the roughness. These variations are due to the densification of the metal film upon crystallization and to the formation of crystallites modifying the spatial distribution of the roughness.

From the methodological point of view, our approach consisting in obtaining non-destructively geometrical parameters (thickness and roughness height) from XRR in the hard x-ray range and chemical composition from XES and then validating this description by XRR in the soft x-ray range appears to be efficient. The methodology can be applied provided that the extrapolation of the parameters from the hard to the soft x-ray range is possible.

***Acknowledgments****: A part of this work was supported by the European Community - Research Infrastructure Action under the FP6 "Structuring the European Research Area" Programme (through the Integrated Infrastructure Initiative "Integrating Activity on Synchrotron and Free Electron Laser Science") under the contract RII3-CT-2004-506008 (IA-SFS). All multilayer depositions have been carried out on the deposition machine of CEMOX (Centrale d'élaboration et de métrologie des optiques X) implemented at the Institut d'Optique by PRaXO (Pôle d'optique des Rayons X d'Orsay).*

**References**

[1] P. Jonnard, I. Jarrige, R. Benbalagh, H. Maury, J.-M. André, Z. Dankhazi, G. Rolland, Surface Science 589 (2005) 164-172.

[2] H. Maury, J.-M. André, J. Gautier, F. Bridou, F. Delmotte, M.F. Ravet, P. Holliger, P. Jonnard, Surface and Interface Analysis 38 (2006) 744-747.

[3] H. Maury, P. Jonnard, J.-M. André, J. Gautier, F. Bridou, F. Delmotte, M.F. Ravet, Surface Science 601 (2007) 2315-2322.

[4] H. Maury, PhD Thesis, Université Pierre et Marie Curie, Paris, 2007.

[5] H. Maury, P. Jonnard, J.-M. André, J. Gautier, M. Roulliay, F. Bridou, F. Delmotte, M.F. Ravet, A. Jérome, P. Holliger, Thin Solid Films 514 (2006) 278-286.

[6] H. Maury, P. Jonnard, K. Le Guen, J.-M. André, Z. Wang, J. Zhu, J. Dong, Z. Zhang, F. Bridou, F. Delmotte, C. Hecquet, N. Mahne, A. Giglia, S. Nannaronne, European Physical Journal B**64** (2008) 193-199.

[7] J. Gautier, F. Delmotte, M. Roulliay, F. Bridou, M.F. Ravet, A. Jérome, Applied Optics 44 (2005) 384-390.

[8] P. Jonnard, C. Bonnelle, A. Bossebœuf, K. Danaie, E. Beauprez, Surface and Interface Analysis 29 (2000) 255-259.

[9] I. Jarrige, P. Jonnard, N. Frantz-Rodriguez, K. Danaie, A. Bossebœuf, Surface and Interface Analysis 34 (2002) 694-697.

[10] C. Bonnelle, F. Vergand, P. Jonnard, J.-M. André, P.F. Staub, P. Avila, P. Chargelègue, M.F. Fontaine, D. Laporte, P. Paquier, A. Ringuenet, B. Rodriguez, Review of Scientific Instruments 65 (1994) 3466-3471.

[11] L. Névot, B. Pardo, J. Corno, Revue de Physique Appliquee 23 (1988) 1675-1686.

[12] S. Nannarone, F. Borgatti, A. DeLuisa, B.P. Doyle, G.C. Gazzadi, A. Giglia, P. Finetti, N. Mahne, L. Pasquali, M. Pedio, G. Selvaggi, G. Naletto, M.G. Pelizzo, G. Tondello, AIP Conference Proceedings (2004) 450-453.

[13] U. Pietsch, V. Holy, T. Baumbach, High-Resolution X-Ray Scattering, From Thin Films, to Lateral Nanostructures, second ed., Springer-Verlag, New-York, 2004.

[14] N. Miyata, S. Ishikawa, M. Yanagihara, M. Watanabe, Japanese Journal of Applied Physics, Part 1 (Regular Papers, Short Notes & Review Papers) 38 (1999) 6476-6478.

[15] C. Largeron, E. Quesnel, J. Thibault, Philosophical Magazine 86 (2006) 2865-2879.

[16] I. Nedelcu, R.W.E. van de Kruijs, A.E. Yakshin, F. Tichelaar, E. Zoethout, E. Louis, H. Enkisch, S. Muellender, F. Bijkerk, Thin Solid Films 515 (2006) 434-438 ?

[17] M. Rovezzi, F. D'Acapito, A. Patelli, V. Rigato, G. Salmaso, E. Bontempi, I. Davoli, Nuclear Instruments & Methods in Physics Research, Section B (Beam Interactions with Materials and Atoms) 246 (2006) 127-130.

[18] F. Bridou, B.A. Pardo, Journal of Optics 21 (1990) 183-191.

[19] J. Shen, A.A. Maradudin, Physical Review B 22 (1980) 4234-4240.

[20] J.-M. André, Optics Communications 52 (1984) 87-93.

[21] R.M. Fitzgerald, A.A. Maradudin, Waves in Random Media 4 (1994) 275-296.

[22] K. Holloway, D. Khiem Ba, R. Sinclair, Journal of Applied Physics 65 (1989) 474-480.

[23] D.G. Stearns, R.S. Rosen, S.P. Vernon, Journal of Vacuum Science & Technology A (Vacuum, Surfaces, and Films) 9 (1991) 2662-2669.

[24] J.M. Slaughter, D.W. Schulze, C.R. Hills, A. Mirone, R. Stalio, R.N. Watts, C. Tarrio, T.B. Lucatorto, M. Krumrey, P. Mueller, C.M. Falco, Journal of Applied Physics 76 (1994) 2144-2156.

[25] D.P. Adams, L.J. Parfitt, J.C. Bilello, S.M. Yalisove, Z.U. Rek, Thin Solid Films 266 (1995) 52-57.

[26] S. Bajt, D.G. Stearns, P.A. Kearney, Journal of Applied Physics 90 (2001) 1017-1025.

[27] S. Yulin, T. Feigl, T. Kuhlmann, N. Kaiser, A.I. Fedorenko, V.V. Kondratenko, O.V. Poltseva, V.A. Sevryukova, A.Y. Zolotaryov, E.N. Zubarev, Journal of Applied Physics 92 (2002) 1216-1220.

[28] M. Nayak, G.S. Lodha, R.V. Nandedkar, S.M. Chaudhari, P. Bhatt, Journal of Electron Spectroscopy and Related Phenomena 152 (2006) 115-120.

[29] D.L. Windt, Computers in Physics 12 (1998) 360-370.

**Figure captions**

Figure 1: Si Kβ soft x-ray emission spectra of the sample Mo-2 and the various references and fit of the multilayer spectrum by a weighted sum of the reference spectra.

Figure 2: X-ray reflectivity of the Mo-2 sample obtained at 0.154 nm with its fit by a model with four layers in a period.

Figure 3: Reflectivity of the Mo-2 multilayer at 1.33 nm (dots) with the simulation (line) deduced from XES and XRR at 0.154 nm.

Figure 4: Transverse scans at 1.33 nm around the first (a) and second (b) Bragg peaks for multilayers Mo-2 (dashed line), Mo-3 (dotted line) and Mo-4 (solid line).

Figure 5: Transverse scan (dots) at 1.33 nm around the first Bragg peak of the Mo-2 sample and simulation (line) with $\xi = 20$ nm and $L = 10$ nm.

Figure 6: Transverse scan (dots) at 1.33 nm around the second Bragg peak of the Mo-2 sample and simulation with $\xi = 20$ nm (solid line) et $\xi = 18$ nm (dotted line) and $L = 10$ nm.

**Table captions**

Table I: Name of the Mo/Si multilayers and expected thickness of the Mo layers from single layer calibration of the deposition process. All the Si layers are 5 nm thick.

Table II: Thickness and roughness of the various layers of the Mo/Si multilayers deduced from XES and from the fit of the reflectivity curves obtained at 0.154 nm. IL1 means interlayer at the Mo-on-Si interface. IL2 means interlayer at the Si-on-Mo interface. $\delta$ is the unit decrement of the real part of the refractive index.

Table III: Lateral correlation lengths $\xi$ of the roughness (nm) of the Mo/Si multilayers.